\documentstyle[twoside,fleqn,espcrc2]{article}

\begin{document}

\def\a{\alpha}
\def\b{\beta}
\def\c{\gamma}
\def\d{\delta}
\def\e{\epsilon}
\def\h{\eta}
\def\k{\kappa}
\def\l{\lambda}
\def\m{\mu}
\def\n{\nu}
\def\o{\theta}
\def\p{\pi}
\def\r{\rho}
\def\s{\sigma}
\def\t{\tau}
\def\u{\upsilon}
\def\w{\omega}
\def\x{\chi}
\def\y{xi}
\def\z{\zeta}
 
\def\C{\Gamma}
\def\D{\Delta}
\def\L{\Lambda}
\def\O{\Theta}
\def\P{\Pi}
\def\S{\Sigma}
\def\W{\Omega}

\def\pl{\partial}
\def\rta{\rightarrow}
\def\la{\langle}
\def\ra{\rangle}
\def\DD{{\cal D}}
\def\CbA{\C_{\phi_5^b A^\l A^\r}}
\def\CcA{\C_{\phi_5^c A^\l A^\r}}
\def\CQA{\C_{Q A^\l A^\r}}
\def\CnA{\C_{\eta^\a A^\l A^\r}}
\def\CGA{\C_{G A^\l A^\r}}
\def\epp{\e_{\l\r\a\b}p_1^\a p_2^\b}

\newcommand{\beq}{\begin{equation}}
\newcommand{\eeq}{\end{equation}}
\newcommand{\beqa}{\begin{eqnarray}}
\newcommand{\eeqa}{\end{eqnarray}}

\newcommand{\ttbs}{\char'134}
\newcommand{\AmS}{{\protect\the\textfont2
  A\kern-.1667em\lower.5ex\hbox{M}\kern-.125emS}}

\title{$\eta'(\eta)\rightarrow\gamma\gamma$ : A Tale of Two Anomalies}

\author{G.M. Shore \address{Department of Physics,\\
        University of Wales, Swansea,\\
        Swansea SA2 8PP, United Kingdom,\\
        {\it and}\\
        Laboratoire de Physique Math\'ematique et Th\'eorique,\\
        Universit\'e de Montpellier II,\\
        F-34095 Montpellier Cedex 05, France.\\}
\thanks{Review article prepared for the `Workshop on Eta Physics', 
University of Uppsala, October 2001. Report no: SWAT/324.}}

\begin{abstract}
The radiative decays of the pseudoscalar mesons $P=(\pi^0,\eta,\eta')$
are of special interest as they provide an experimental window on the 
electromagnetic and colour chiral $U_A(1)$ anomalies. 
$\pi^0\rightarrow\gamma\gamma$ is well described 
by the electromagnetic $U_A(1)$ anomaly under the assumption that $\pi^0$ 
is a Goldstone boson for spontaneously broken chiral symmetry, but the
analogous results for $\eta'(\eta)\rightarrow\gamma\gamma$ are complicated 
by the colour $U_A(1)$ anomaly in QCD. This paper reviews the theory 
of $\eta'(\eta)\rightarrow\gamma\gamma$ decays, emphasising the role of the
colour $U_A(1)$ anomaly and the renormalisation group. The relation
to the Witten-Veneziano mass formula for the $\eta'$ and the QCD 
topological susceptibility is derived. The implications for the
phenomenological analysis of $P\rightarrow\gamma\gamma$ are described
and a proposal is made for a comprehensive re-analysis of existing data
on $P\rightarrow\gamma\gamma$, $P\rightarrow V\gamma$, where $V=(\rho,\omega,
\phi)$ is a light $1^-$ meson, $\eta'(\eta)\rightarrow \pi\pi\gamma$, 
$\psi\rightarrow\eta'(\eta)\gamma$, etc. 
Other applications, to the muon anomalous magnetic moment $g_\m-2$ and
the determination of the photon structure function $g_1^\gamma$ in
polarised deep inelastic scattering, are also briefly discussed.

\end{abstract}
 
\maketitle
 
\section{Principal Results}

The radiative decays of the pseudoscalar mesons $P\rta\c\c$, where
$P=(\pi^0,\eta,\eta')$ have long been of special interest because of their
intimate relation with the chiral (axial) $U_A(1)$ anomaly. The decay
$\pi^0\rta\c\c$ is of course the textbook example of the phenomenological
importance of the electromagnetic contribution to the $U_A(1)$ anomaly.
It has played an important role both in understanding the theory of
anomalies and in pinning down the quantum numbers of the quarks and
number of colours $N_c$ in the early days of QCD. 
The decay $\eta'\rta\c\c$ is doubly interesting, since it also involves in 
an essential way the colour contribution to the $U_A(1)$ anomaly.

An immediate consequence of the colour $U_A(1)$ anomaly in QCD is that even 
in the limit of massless quarks, the $\eta'$, unlike the $\pi^0$ and $\eta$, 
is not a Goldstone boson of spontaneously broken chiral symmetry. 
The success of the theoretical prediction relating the decay amplitude for 
$\pi^0\rta\c\c$ to the electromagnetic $U_A(1)$ anomaly is based on the pion 
being an approximate Goldstone boson in the massless limit. The complete
theoretical analysis can then be abbreviated in `soundbite' form by 
writing the familiar PCAC relation:
\beq
\pl^\m J_{\m5}^3 \rta f_\pi m_\pi^2 \pi
\eeq
where onthe r.h.s. $\pi$ is a phenomenological field for the $\pi^0$.
The corresponding analysis for the $\eta'$ is complicated first by the
importance of non-trivial $\eta - \eta'$ mixing due to the explicit
$SU(3)$ breaking induced by $m_s \gg m_u,m_d$ and second, by the presence of
the colour contribution to the $U_A(1)$ anomaly. Nevertheless, the analogous
analysis can be pushed through using the anomalous Ward identities and, 
subject to reasonable dynamical assumptions, results in the following
generalised PCAC relation for the $\eta$ and $\eta'$:
\beq
\pl^\m J_{\m5}^a \rta f^{a\a} m_{\a\b}^2 \eta^\b + 2n_f G \d_{a0}
\eeq
where $\eta^\a = (\eta,\eta')$ and $a=0,8$. 
The new term on the r.h.s. reflects the presence of the anomaly. 
Very roughly, what happens is that due to the anomaly, the flavour singlet 
pseudoscalar mixes with the gluonic operator $G^{\m\n}\tilde G_{\m\n}$. 
The conventional PCAC relation applies only to the fictitious state before 
mixing. Rearranging this into a part involving the physical $\eta'$
therefore leaves a residual `glue' contribution, represented by $G$. 
However, great care must be taken with this picture and especially with
the interpretation of eq.(2). In particular, $G$ is {\it not} to be understood
necessarily as the lightest $0^-$ glueball, and the relation (2)
is only valid as an operator relation for insertions into 
{\it zero-momentum} Green functions. These points will be explained 
carefully in due course. 

In this paper we review the special features that arise 
due to the gluonic contribution to the $U_A(1)$ anomaly when PCAC
methods (which include chiral Lagrangians) are used to analyse radiative 
pseudoscalar meson decays in the flavour singlet
channel. In \cite{SVeta}, we presented
an analysis of $\eta' \rta \c\c$ decay in the chiral limit
of QCD, taking into account the gluonic anomaly and the associated anomalous 
scaling implied by the renormalisation group. Here, we summarise the results
of a more recent analysis\cite{Seta} extending this 
to QCD with massive quarks, incorporating $\eta - \eta'$ mixing. In
particular, we show how a combination of the radiative decay formula
and a generalisation of the Witten--Veneziano mass formula for the $\eta'$ 
could be used to measure the gluon topological
susceptibility $\chi(0)$ in full QCD with massive quarks.

Our main result is summarised in the formulae:
\beq
f^{a\a}~ g_{\eta^\a\c\c} ~+~ 2n_f A~ g_{G\c\c}~ \d_{a0} ~=~ 
a_{\rm em}^a {\a\over\pi}
\eeq
which describes the radiative decays, and
\beqa
&f^{a\a} (m^2)_{\a\b} f^{T\b b} = (2n_f)^2 A~\d_{a0}\d_{b0}\cr 
{}\cr 
&-2 d_{abc} {\rm tr}~ T^c
\left(\matrix{m_u\la\bar u u\ra &0 &0 \cr
0 &m_d\la\bar d d\ra &0 \cr 
0 &0 &m_s\la\bar s s\ra } \right) \cr
\eeqa
which defines the decay constants appearing in (3) through a 
modification of the Dashen, Gell-Mann Oakes Renner \footnote{
Hereafter simply referred to as the Dashen formula.} \cite{GMOR,Dash}
formula to include the gluon contribution to 
the $U_A(1)$ anomaly. 

In these formulae, $\eta^\a$ denotes the neutral pseudoscalars 
$\pi^0,\eta,\eta'$. The (diagonal) mass matrix is $(m^2)_{\a\b}$ 
and $g_{{\eta^\a}\c\c}$ is the appropriate coupling, 
defined as usual from the decay amplitude by
$\la \c\c |\eta^\a \ra = -i g_{\eta^\a\c\c} \e_{\l\r\a\b}p_1^\a p_2^\b
\e^\l(p_1) \e^\r(p_2)$. 
The constant $a_{\rm em}^a$ is the coefficient of the
electromagnetic contribution to the axial current anomaly:
\beq
\pl^\m J_{\m5}^a = d_{acb} m^c \phi_5^b  +
2n_f \d_{a0} Q  + a_{\rm em}^a {\a\over8\pi} F^{\m\n}\tilde F_{\m\n} 
\eeq
Here, $J_{\m5}^a$ is the axial current, $\phi_5^a = \bar q \c_5 T^a q$
is the quark pseudoscalar and $Q = {\a_s\over8\pi}{\rm tr}G^{\m\n} 
\tilde G_{\m\n}$ is the gluon topological charge. 
$m^a$ are the quark masses (see eq.(16)). $a=0,3,8$ is the flavour 
index, $T^{3,8}$ are $SU(3)$ generators and $T^0 = {\bf 1}$.
The $d$-symbols are defined by $\{T^a,T^b\} = d_{abc} T^c$. Since this
includes the flavour singlet $U_A(1)$ generator, they are only symmetric on
the first two indices. For $n_f = 3$, the explicit values are $d_{000} =
d_{033} = d_{088} = 2, d_{330} = d_{880} = 1/3, d_{338} = d_{383} =
-d_{888} = 1/\sqrt3$. 

These formulae become clearer if we immediately make the approximation 
$m_u,m_d \ll m_s$. The formulae for the $\pi^0$ decouple:
\beq
f_\pi g_{\pi\c\c} = a_{\rm em}^3 {\a\over\pi}
\eeq
where $a_{\rm em}^3= {1\over3}N_c$ and
\beq
f_\pi^2 m_\pi^2 = - (m_u \la\bar u u\ra + m_d \la\bar d d\ra)
\eeq
leaving the following new results\cite{Seta} for the $\eta$ and $\eta'$:
\beqa
{}&f^{0\eta'} g_{\eta'\c\c} + f^{0\eta} g_{\eta\c\c}
+ 6 A g_{G\c\c} = a_{\rm em}^0 {\a\over\pi} ~~~~~~\cr
{}\cr
{}&f^{8\eta} g_{\eta\c\c} + f^{8\eta'} g_{\eta'\c\c}
= a_{\rm em}^8 {\a\over\pi} \cr
\eeqa
where $a_{\rm em}^0 = {4\over3} N_c$ and $a_{\rm em}^8 = {1\over3\sqrt3} N_c$,
together with the generalised Dashen formulae:
\beqa
&\bigl(f^{0\eta'}\bigr)^2 m_{\eta'}^2 +
\bigl(f^{0\eta}\bigr)^2 m_{\eta}^2 = -4m_s \la\bar s s\ra + 36 A \cr
{}\cr
&f^{0\eta'} f^{8\eta'} m_{\eta'}^2 +
f^{0\eta} f^{8\eta} m_{\eta}^2 = {4\over\sqrt3} m_s \la\bar s s\ra \cr
{}\cr
&\bigl(f^{8\eta}\bigr)^2 m_{\eta}^2 +
\bigl(f^{8\eta'}\bigr)^2 m_{\eta'}^2 = -{4\over3}m_s \la\bar s s\ra  \cr
\eeqa

The decay constants $f^{a\a}$ in (3) are defined so as to satisfy the relation (4).
It is crucial to recognise that in general they are {\it not} the couplings 
of the pseudoscalar mesons to the axial current\cite{SVeta}. In the flavour 
singlet sector, such a definition would give a RG non-invariant decay 
constant which would not coincide with those in the correct 
decay formula (3). In contrast, {\it all} the quantities in the formulae
(3),(4) are separately RG invariant\cite{Seta},\cite{Szuoz}. The proof is not 
immediately obvious, and depends on the RGEs for the various Green functions
and vertices defining the terms in (3),(4) being evaluated on-shell or at
zero-momentum.

As we have seen, since flavour $SU(2)$ symmetry is almost exact, the relations
for $\pi^0$ decouple and are simply the standard ones with $f^{3\pi}$
identified as $f_\pi$.
In the octet-singlet sector, however, there is mixing and the decay constants
form a $2\times 2$ matrix:
\beq
f^{a\a} = \left(\matrix{f^{0\eta'} & f^{0\eta} \cr
f^{8\eta'} & f^{8\eta}} \right) 
\eeq
The four components are independent. In particular, for broken $SU(3)$,
there is no reason to express $f^{a\a}$ as a diagonal matrix times an
orthogonal $\eta - \eta'$ mixing matrix, which would give just three
parameters. Several convenient parametrisations may be made, 
e.g.~involving two constants and two mixing angles, but this does not  
reflect any special dynamics. We return to this point when we discuss
phenomenology in section 3.

The novelty of our results of course lies in the extra terms arising in
(3) and (4) due to the gluonic contribution to the $U_A(1)$ anomaly.
The coefficient $A$ is the non-perturbative number which specifies the 
topological susceptibility in full QCD with massive dynamical quarks.
The topological susceptibility is defined as
\beq
\chi(0) = \int d^4x~i\la0|T~Q(x)~Q(0)|0\ra
\eeq
The anomalous chiral Ward identities determine 
its dependence on the quark masses and condensates up to an 
undetermined parameter, viz.
\beq
\chi(0) = -A \biggl(1 - A \sum_q {1\over m_q \la \bar q q\ra}\biggr)^{-1} 
\eeq
Notice how this satisfies the well-known result that $\chi(0)$ vanishes
if any quark mass is set to zero.

The modified flavour singlet Dashen formula is in fact a generalisation 
of the Witten--Veneziano mass formula for the $\eta'$. 
Here, however, we do {\it not} impose the leading order in $1/N_c$ 
approximation that produces the Witten--Veneziano formula. 
Recall that this states
\beq
m_{\eta'}^2 + m_{\eta}^2 - 2 m_K^2 =    
-{6\over f_\pi^2} \chi(0)\big|_{\rm YM}
\eeq
To recover (13) from our result (see the first of eqs.(9)) the condensate 
$m_s \la \bar s s\ra$ is replaced by the term proportional to $f_\pi^2 m_K^2$ 
using a standard Dashen equation, and the singlet decay constants are set to 
$\sqrt{2n_f}f_\pi$.
The identification of the large $N_c$ limit of the coefficient $A$ with
the non-zero topological susceptibility of pure Yang-Mills theory follows
from large $N_c$ counting rules and is explained in ref.\cite{Seta}.
Lattice evaluations \cite{Pisa} (see also ref.\cite{Rossi}) in the quenched 
approximation find the following value,
\beq
\chi(0)\Big|_{\rm YM} = - (180 {\rm MeV})^4
\eeq
A similar result has also been obtained using QCD spectral sum rules
\cite{Nar}.

The final element in (3) is the extra `coupling' $g_{G\c\c}$
in the flavour singlet decay formula, which arises because even in the 
chiral limit the $\eta'$ is not a Goldstone boson because of the gluonic
$U_A(1)$ anomaly. A priori, this is {\it not} the coupling of a physical
particle, although (suitably normalised) it could be modelled as the coupling
of the lightest predominantly glueball state mixing with $\eta'$. However,
this interpretation would probably stretch the basic dynamical 
assumptions underlying (3) too far, and is not necessary
either in deriving or interpreting the formula. In fact, the $g_{G\c\c}$
term arises simply because in addition to the electromagnetic anomaly,
the divergence of the axial current contains both the quark pseudoscalar  
$\phi_5^a$ and the gluonic anomaly $Q$. 
Diagonalising the propagator matrix for these operators isolates the
$\eta$ and $\eta'$ poles, whose couplings to $\c\c$ give the usual
terms $g_{\eta\c\c}$ and $g_{\eta'\c\c}$. However, the remaining
operator (which we call $G$) also couples
to $\c\c$ and therefore also contributes to the decay formula, whether
or not we assume that its propagator is dominated by a `glueball' pole.
We emphasise again that there is no need whatsoever to make any assumption
about the spectrum of the $G$ propagator in deriving the decay formulae. 

The presence of the coupling $g_{G\c\c}$ in (8) however removes any 
immediate predictivity from the $\eta'\rta\c\c$ decay formula.
In order to push the theory further, we therefore need to make additional
dynamical assumptions. The obvious possibility is to explore whether
the phenomenologically successful OZI rule can be applied in this context.
(The OZI rule is of course closely related (but not identical\cite{Vozi}) to the
$1/N_c$ expansion, which is discussed in this context in ref.~\cite{SVeta}.)
In fact, as we now argue, the OZI rule does provide some justification for
believing that the extra coupling $g_{G\c\c}$ is indeed small.
The theoretical argument is based on the fact that $g_{G\c\c}$ 
is both OZI suppressed {\it and} renormalisation group (RG) 
invariant\cite{SVeta}. Since violations of the OZI rule are
associated with the $U_A(1)$ anomaly, it is a plausible conjecture
that we can identify OZI-violating quantities by their dependence on the 
anomalous dimension associated with the non-trivial renormalisation of 
$J_{\m5}^0$ due to the anomaly. In this way, RG non-invariance can be used 
as a flag to indicate those quantities expected to show large OZI violations.
If this conjecture is correct, then we would expect the OZI rule to be
reasonably good for the RG invariant $g_{G\c\c}$, which would therefore
be suppressed relative to $g_{\eta'\c\c}$.
(An important exception to this argument is of course the $\eta'$ mass itself, 
which although obviously RG invariant is not zero in the chiral limit 
as it would be in the OZI limit of QCD.) Notice 
that this conjecture has been applied already with some success to the
`proton spin' problem in polarised deep inelastic scattering\cite{Serice}.

If it is indeed a good dynamical approximation to assume $g_{G\c\c}$ is 
small compared to $g_{\eta'\c\c}$, we could combine eqs (8)
and (9) to give a measurement of the non-perturbative coefficient $A$ in
$\chi(0)$. To see this, assume that the physical quantities $m_{\eta}, 
m_{\eta'}, g_{\eta\c\c}$ and $g_{\eta'\c\c}$ are all known and (temporarily) 
neglect $g_{G\c\c}$.
Clearly, the two purely octet formulae can be used to find $f^{8\eta}$
and $f^{8\eta'}$ if both $g_{\eta\c\c}$ and $g_{\eta'\c\c}$ are known.
The off-diagonal Dashen formula then expresses $f^{0\eta}$ in terms of
$f^{0\eta'}$.
This leaves the two purely singlet formulae involving the still-undetermined
decay constant $f^{0\eta'}$, the topological susceptibility coefficient $A$,
and the coupling $g_{G\c\c}$.
The result follows immediately. If we neglect $g_{G\c\c}$,
we can find $f^{0\eta'}$ from the singlet decay formula and thus
determine $A$ from the remaining flavour singlet Dashen formula. 
This is the generalisation of the Witten--Veneziano formula.
Determining $A$ in this way would of course be an important link between the 
phenomenology of $\eta'$ decays and the important subject of gluon topology 
in QCD \cite{Szuoz}.

However, the issue of the magnitude of $g_{G\c\c}$ is ultimately an
experimental question. It is therefore crucial that phenomenological
analyses of the data on $\eta$ and $\eta'$ decays start from the
complete formulae (8),(9) and do a best fit involving the full set
of parameters. Only then will we really know whether or not the extra
anomaly-induced coupling $g_{G\c\c}$ is small.

We return to this issue in section 3. First, we provide a PCAC-based
proof of the key formulae (3),(4). Carefully used, these methods 
permit a very quick derivation. However, there are many subtleties
in the analysis which are easily missed using these conventional techniques.
For this reason, we prefer the more field-theoretic approach explained
in refs.\cite{SVeta,Seta} (see also \cite{Szuoz}). This makes very clear
exactly what dynamical assumptions must be made and where they enter
the argument. The price is working with a formalism which, although
elegant and easy to use, is unfamiliar to many phenomenologists.
We present this field-theoretic derivation in section 5, although readers
interested only in the phenomenological results will find the discussion
in sections 1--4 complete and self-contained.
 
\vskip0.5cm

\section{$U_A(1)$ PCAC}

Consider first QCD by itself without the coupling to electromagnetism.
The axial anomaly is
\beq
\pl^\m J_{\m5}^a = M_{ab}\phi_5^b + 2n_f Q \d_{a0}
\eeq
The notation is defined in ref.\cite{Seta}.
The quark mass matrix is written as $m^a T^a$, so  
\beq
\left(\matrix{m_u &0 &0 \cr
0 &m_d &0 \cr
0 &0 &m_s \cr}\right)
= m^0 {\bf 1} + m^3 T^3 + m^8 T^8
\eeq
The condensates are written as 
\beq
\left(\matrix{ \langle \bar u u\rangle &0 &0 \cr 0 &\langle \bar d d\rangle
&0 \cr
0 &0 &\langle \bar s s\rangle \cr}\right) = {1\over 3} \la\phi^0\ra {\bf 1}
+ 2
\la\phi^3\ra T^3 + 2 \la\phi^8\ra T^8 
\eeq
where $\langle \phi^c\rangle$ is the VEV $\langle \bar q T^c q\rangle$.
Then
\beq
M_{ab} = d_{acb} m^c, ~~~~~~
\Phi_{ab} = d_{abc} \langle \phi^c\rangle 
\eeq

An essential approximation in the application of PCAC methods is to use 
identities valid at zero momentum and make certain smoothness assumptions
(see section 5 for a careful discussion) to extrapolate to the mass-shell
of the physical states. We therefore use the zero-momentum chiral
Ward identities. For the two-point Green functions of the relevant operators 
these are\cite{Seta}
\beqa
&2n_f \la Q~Q\ra \d_{a0} + M_{ac} \la \phi_5^c~Q\ra = 0 \cr
{}\cr
&2n_f \la Q~\phi_5^b\ra \d_{a0} + M_{ac}\la\phi_5^c~\phi_5^b\ra 
+ \Phi_{ab} = 0 \cr
\eeqa
which imply
\beq
M_{ac} M_{bd} \la \phi_5^c~\phi_5^d\ra = - (M\Phi)_{ab} 
+ (2n_f)^2 \la Q~Q\ra \d_{a0}\d_{b0}
\eeq
We also need the result for the general form of the topological 
susceptibility (see eq.(12)):
\beq
\chi(0) \equiv \la Q~Q \ra = {-A\over 1 - (2n_f)^2 A (M\Phi)_{00}^{-1} }
\eeq

Although the pseudoscalar operators $\phi_5^a$ and $Q$ couple to the 
physical states $\eta^\a = (\pi^0,\eta,\eta')$, it is more convenient to
redefine linear combinations such that the resulting propagator matrix
is diagonal and properly normalised. So we define operators
$\eta^\a$ and $G$ such that
\beq
\left(\matrix{\la Q~Q \ra &\la Q~\phi_5^b \ra \cr
\la \phi_5^a ~Q \ra &\la \phi_5^a~\phi_5^b \ra }\right) ~~\rta~~
\left(\matrix{\la G~G\ra &0 \cr
0 &\la \eta^\a ~ \eta^\b \ra }\right)
\eeq
This is achieved by 
\beq
G =  Q - \la Q~\phi_5^a\ra (\la \phi_5 ~\phi_5\ra)_{ab}^{-1} \phi_5^b 
\eeq
which reduces at zero momentum to
\beq
G= Q + 2n_f A \Phi_{0b}^{-1} \phi_5^b
\eeq
and we define
\beq
\eta^\a = f^{T\a a} \Phi_{ab}^{-1} \phi_5^b 
\eeq
With this choice, the $\la G~G\ra$ propagator at zero momentum is
\beq
\la G~G\ra = - A
\eeq
and we impose the normalisation
\beq
\la \eta^\a ~\eta^\b\ra = {-1\over k^2 - m_{\eta^\a}^2}\d^{\a\b}
\eeq
This implies that the constants $f^{a\a}$ in (25), which 
are simply the decay constants, must satisfy the (Dashen) identity
\beqa
f^{a\a} m_{\a\b}^2 f^{T\b b} &= \Phi_{ac} (\la \phi_5~\phi_5 \ra)_{cd}^{-1}
\Phi_{db} ~~~~~~~~~~ \cr
{}\cr
&= -(M\Phi)_{ab} + (2n_f)^2 A \d_{a0}\d_{b0} \cr
\eeqa
The last line follows from the Ward identities (20) and (21).
In terms of these new operators, the anomaly equation (14) at zero
momentum is:
\beq
\pl^\m J_{\m5}^a  \rta f^{a\a} m_{\a\b}^2 \eta^\b + 2n_f G \d_{a0}
\eeq
The notation $\rta$ is to emphasise that the identity is only true
for insertions of the operators into zero-momentum Green functions
and matrix elements. At non-zero momentum, other operators appear
on the r.h.s. (In particular, we can not take the 
{\it on-shell} matrix elements between the vacuum and the $|\eta^\a\ra$ and
use $\la 0|G|\eta^a\ra = 0$ to conclude that the decay constants
$f^{a\a}$ can be identified as usual with the matrix elements 
$\la 0|\pl^\m J_{\m5}^a |\eta^\a\ra$. As we have repeatedly emphasised,
the anomaly removes the familiar correspondence between the decay constants
and the current matrix elements.) 

Now recall how conventional PCAC is
applied to the calculation of $\pi^0\rta\c\c$. The pion decay constant
is defined as the coupling of the pion to the axial current
\beq
\la 0|J_{\m5}^3|\pi\ra = ik_\m f_\pi ~\Rightarrow~
\la 0|\pl^\m J_{\m5}^3|\pi\ra = f_\pi m_\pi^2
\eeq
and satisfies the usual Dashen formula
The next step is to define a `phenomenological pion field' $\pi$ by
\beq
\pl^\m J_{\m5}^3 \rta f_\pi m_\pi^2 \pi
\eeq
To include electromagnetism, the full anomaly equation is extended as in
(5) to include the $F^{\m\n} \tilde F_{\m\n}$ contribution. Using (29)
we have
\beqa
&ik^\m\la \c\c|J_{\m5}^3|0 \ra ~~~~~~~~~~~~~~~~~ \cr
{}\cr
&= f_\pi m_\pi^2 \la \c\c|\pi|0\ra
+ a_{\rm em}^a {\a\over8\pi} \la \c\c|F^{\m\n} 
\tilde F_{\m\n}|0\ra~~~~\cr
{}\cr
&= f_\pi m_\pi^2 \la \pi~\pi\ra \la \c\c|\pi\ra
+ a_{\rm em}^a {\a\over8\pi} \la \c\c|F^{\m\n} \tilde F_{\m\n}|0\ra \cr
\eeqa
where $\la \pi~\pi\ra$ is the pion propagator $-1/(k^2-m_\pi^2)$.
At zero momentum, the l.h.s.~vanishes because of the explicit $k_\m$
factor and the absence of massless poles. We therefore find, 
\beq
f_\pi g_{\pi\c\c} = a_{\rm em}^3 {\a\over\pi}
\eeq

In the full theory including the flavour singlet sector and the gluonic
anomaly, we find a similar result. The `phenomenological fields'
are defined by (29) where the decay constants satisfy the generalised
Dashen formula (28) . Notice again that they are {\it not} simply
related to the couplings to the axial current as in (31) for the
flavour non-singlet. We therefore find:
\beqa
{}&ik^\m\la \c\c|J_{\m5}^a|0 \ra 
= f^{a\a} m_{\a\b}^2 \la \c\c|\eta^\b|0\ra ~~~~~~~~~~~~~~~~\cr
{}&+ 2n_f \la \c\c|G|0\ra \d_{a0} 
+ a_{\rm em}^a {\a\over8\pi} \la \c\c|F^{\m\n} \tilde F_{\m\n}|0\ra \cr
{}&{}\cr
{}&~~~~~~~~~= f^{a\a} m_{\a\b}^2 \la \eta^\b~\eta^\c \ra \la \c\c|\eta^\c \ra \cr
{}&+ 2n_f \la G~G\ra \la \c\c|G\ra \d_{a0} 
+ a_{\rm em}^a {\a\over8\pi} \la \c\c|F^{\m\n} \tilde F_{\m\n}|0\ra \cr
\eeqa
using the fact that the propagators are diagonal in the basis $\eta^\a, G$.
Using the explicit expressions (26),(27) for the Green functions,
we find in this case:
\beq
f^{a\a} ~g_{\eta^a\c\c} + 2n_f A ~g_{G\c\c} ~\d_{a0} ~=~ a_{\rm em}^a 
{\a\over\pi}
\eeq
where the extra coupling $g_{G\c\c}$ is defined through (34).
This completes the derivation. It is evidently a
straightforward generalisation of conventional PCAC with the 
necessary modification of the usual formulae to take account of the
extra gluonic contribution to the axial anomaly in the flavour
singlet channel, the key point being the identification of the 
operators $\eta^\a$ and $G$ in (29).

\section{Phenomenology}

In this section, we discuss critically the way in which data on 
$\eta'(\eta)\rta\c\c$ decays (and closely related processes such as
$\eta'(\eta)\rta V\c$) are analysed in the phenomenological 
literature\footnote{Here, and in section 4, we have only
cited the few papers which have been most influential in the
preparation of this article. This is not intended to be a comprehensive
review of the diverse and interesting literature on the phenomenology of
$\eta$ and $\eta'$ physics.}.
The existing analyses are based on formulae in which the impact of
the colour $U_A(1)$ anomaly has not been correctly taken into account.
We therefore propose a comprehensive re-analysis of the data based on the
formulae (8),(9) derived above. 

The two-photon decay widths are given by
\beq
\Gamma(\eta'(\eta)\rta\c\c) ~=~ {m_{\eta'(\eta)}^2\over64\pi} 
|g_{\eta'(\eta)\c\c}|^2
\eeq
The current experimental data, quoted in the Particle Data Group tables 
\cite{PDG}
are
\beq
\Gamma(\eta\rta\c\c) ~=~ 0.510 \pm 0.026 ~{\rm keV}
\eeq
arising principally from the 1988 Crystal Ball \cite{CB} and 1990 ASP 
\cite{ASP} results on $e^+ e^- \rta e^+ e^- \eta$ 
(here, we follow the note in the 1994 PDG compilation \cite{PDG94}
and use only the two-photon $\eta$ production data),
and
\beq
\Gamma(\eta'\rta\c\c) ~=~ 4.28 \pm 0.19 ~{\rm keV}
\eeq
dominated by the 1998 L3 data \cite{L3} on the two-photon formation
of $\eta'$ in $e^+ e^- \rta e^+ e^- \pi^+ \pi^- \c$.

For the purposes of comparing our theoretical results with the standard
phenomenology, we focus on the very thorough and complete discussion
in ref.\cite{BFT}, updated in \cite{EF}. These authors have studied not only
the two-photon decays of $\eta'$ and $\eta$ but also the related 
radiative vector-meson decays $\eta'(\eta)\rta V\c$, where 
$V=(\rho,\omega,\phi)$. They also analyse $\psi\rta\eta'(\eta)\c$ decays.

There are two main points of difference between our theory and the analysis
in these papers (and all other phenomenological treatments of which we are 
aware, including the notes on the $\eta'$ and $\eta$ decay constants and 
two-photon decay formulae in the Particle Data Group tables \cite{PDG}). The 
first is that the term involving the coupling $g_G\c\c$ in the decay formula
(8) is neglected. The second is that the decay constants entering the decay 
formulae are assumed to be given (just as for the pion) by the matrix elements
of the axial current.  Further, in \cite{BFT} it is assumed that the decay 
constant matrix (10) has just three independent components, although this is 
generalised to four in the second paper \cite{EF}. 

To make this concrete, let us define quantities $\hat f^{a\a}$ by
\beq
\la0|\pl^\m J_{\m5}^a|\eta^\b\ra ~=~ \hat f^{a\a} m^2_{\a\b}             
\eeq
The decay formulae used in \cite{BFT,EF} are then simply
\beq
\hat f^{a\a} g_{\eta^\a \c\c} ~{\mathop=^{?}}~ a^a_{\rm em}{\a\over\pi}
\eeq
in our notation. The most obvious problem with eq.(40) is that it is
not consistent with the renormalisation group. Since the singlet axial
current $J_{\m5}^0$ is not RG invariant because of the anomaly (see section 5)
neither are the $\hat f^{0\a}$. This is recognised in \cite{BFT,EF} but the 
associated running of the `decay constants' $\hat f^{0\a}$ is assumed there to
be negligible. However, this is not the main problem, which is that the RG
non-invariance of $\hat f^{0\a}$ is a signal that the basic decay formula
(40) is wrong -- as we have seen above, it must be modified to include the
coupling $g_{G\c\c}$. 

The interesting question for phenomenology to address is whether $g_{G\c\c}$
is in fact small, as suggested by the OZI-based discussion above. The success 
of the programme carried through in \cite{BFT,EF} suggests that this may well
be true. However, the only real way to settle the question is to repeat the
analysis using the correct formulae (8) and to fit the data with the full
set of decay constants $f^{a\a}$ (subject to the Dashen constraints (9))
and the extra coupling $g_{G\c\c}$. The actual value of $g_{G\c\c}$ may also
be of relevance in other related processes (see below). 

The second main problem stems from the mistaken identification of the
relevant decay constants with the axial current matrix elements (39). 
It is clear that the anomaly equation may be used to relate the
(non RG invariant) matrix elements $\la 0|Q|\eta^\a\ra$ of the topological 
charge density to the `decay constants' $\hat f^{a\a}$. Specifically,
if we neglect $m_u,m_d$, we have in our notation (c.f.\cite{BFT,EF})
\beq
\la0|Q|\eta^\b\ra ~=~ {1\over6} \Bigl(\hat f^{0\a}m^2_{\a\b} + \sqrt{3}
\hat f^{8\a}m^2_{\a\b} \Bigr)
\eeq
This expresses the ratio of the coupling of 
$Q = {\a_s\over8\pi}{\rm tr}G^{\m\n} \tilde G_{\m\n}$ to $\eta'$ and $\eta$
in terms of the $\hat f^{0\eta'(\eta)}$.
It has been pointed out in \cite{Nov} that the matrix elements 
$\la0|Q|\eta'(\eta)\ra$ enter into the formulae for the radiative decays
$\psi\rta \eta'(\eta)\c$. 
The current data \cite{PDG} is
\beq
\Gamma(\psi\rta\eta\c) / \Gamma_{\rm total}= (0.86 \pm 0.08)\times 10^{-3}
\eeq
and
\beq
\Gamma(\psi\rta\eta'\c) / \Gamma_{\rm total} ~=~ (4.31 \pm 0.30)\times 10^{-3}
\eeq
where
\beq
\Gamma_{\rm total} ~=~ (87 \pm 5)~{\rm keV}
\eeq
Studying these decays therefore gives
information on the `glue content' of the $\eta'$ and $\eta$.
However, in refs.\cite{BFT,EF} the formula (41) is used in conjunction
with the erroneous (40) to relate the  $\la0|Q|\eta'(\eta)\ra$ to the
decay constants appearing in $\eta'(\eta)\rta\c\c$. This is formally incorrect, 
and in any case it must surely be inconsistent to use the naive decay formula 
(40), which is derived on the basis that both $\eta$ and $\eta'$ are true
pseudo-Goldstone bosons, to estimate their anomaly-induced glue content.
In truth, as originally pointed out in \cite{SVeta}, the physical,
RG invariant, decay constants $f^{a\a}$ appearing in the true decay formulae 
for the $\eta'$ and $\eta$ are in principle quite independent of the quantities 
$\hat f^{a\a}$ related to the axial current matrix elements.

A final comment on the theory concerns $\eta - \eta'$ mixing and the
decay constants $f^{a\a}$. For some time, it was usual to
describe $\eta'(\eta)\rta\c\c$ decays in terms of two decay constants ($f^0$ and
$f^8$) and a single $\eta -\eta'$ mixing angle defined in terms of the
$SU(3)$ eigenstates by
\beqa
|\eta'\ra ~~~{\mathop=^{?}}&\cos\vartheta |\eta_0\ra + 
\sin\vartheta |\eta_8\ra ~~~~~~~~~~~~~~~\cr
|\eta\ra ~~~{\mathop=^{?}}&-\sin\vartheta |\eta_0\ra + 
\cos\vartheta |\eta_8\ra ~~~~~~~~~~~~~~~\cr
\eeqa
More recently \cite{Feld1,Feld2,KL1,KL2,EF} etc., it has been realised that the
$f^{a\a}$ decay constant matrix (10) should instead have {\it four} independent
components, which are conventionally parametrised in terms of the
two $f^{0,8}$ and ${\it two}$ angles. In the version favoured by \cite{EF},
these would be defined through
\beq
\left(\matrix{f^{0\eta'}&f^{0\eta}\cr f^{8\eta'}&f^{8\eta}}\right)~=~ 
\left(\matrix{f^0 \cos\vartheta_{\eta'}
&-f^0\sin\vartheta_{\eta}\cr
f^8 \sin\vartheta_{\eta'}&f^8\cos\vartheta_{\eta}}\right) 
\eeq
Results for $\vartheta_{\eta'}$ and $\vartheta_{\eta}$ derived using (40) are 
quoted in ref.\cite{EF}, where the difference between the two angles is claimed 
to be an `energy dependence' of the $\eta -\eta'$ mixing angle, with one
being evaluated on the $\eta'$ mass-shell and the other on the 
$\eta$ mass-shell.
To us, however, it seems that this interpretation has no consequence. It
is clearly correct to use four parameters to describe the decay constant
matrix $f^{a\a}$ and a parametrisation in terms of two constants and
two angles is as good as any other, but we see no reason to superpose the 
notion of an `energy-dependent mixing angle' on this simple formalism.

The analysis of $\eta'(\eta)\rta\c\c$ decays presented here can 
clearly be extended to study $\eta'(\eta)\rta V\c$, where $V$ is a 
flavour-singlet vector meson $\rho,\omega,\phi$, using the familiar 
assumptions of vector meson dominance. This allows us to extract the couplings 
$g_{\eta'(\eta)V\c}$ from the 3-point Green functions $\la 0|J_{\m5}^{0,8}~ 
J_{\n}^{\rm em}~ J_{\l}^V|0\ra$ by relating them to the 
appropriate electromagnetic anomaly coefficients in exactly the same way as 
for the two-photon decays, then assuming pole-dominance for the vector current.
This is explained in detail in ref.\cite{BFT}. However, exactly the same
comments apply. The true decay formulae for $P\rta V\c$ involve extra
$g_{GV\c}$ couplings in the singlet sector (the full set of formulae
analogous to those quoted in ref.\cite{BFT} can be trivially derived
using the method described in section 2). These are omitted in ref.\cite{BFT}.
Again, the apparently reasonable agreement with data \cite{PDG} found there
suggests that these new couplings may be small, but this should be checked
by repeating the analysis including the $g_{GV\c}$ and finding the best fit
to data. It would also be interesting to make a detailed comparison with 
the quark model/OZI analysis of ref.\cite{BES}.

Another obvious extension is to the decays $\eta'(\eta)\rta\pi^+\pi^-\c$,
related to the 4-point Green functions $\la0|J_{\m5}^{0,8}~ 
J_{\n5}^+ ~ J_{\l5}^- ~J_{\r}^{\rm em}|0\ra$. This Green function has
an AAAV box anomaly in direct analogy to the AVV triangle anomaly
considered above. Clearly this can be analysed simply using the 
$U_A(1)$ PCAC theoretical methods described here. On the experimental side,
while $\eta\rta\pi^+\pi^-\c$ has been observed with decay width\cite{PDG}
\beq
\Gamma(\eta\rta\pi^+\pi^-\c) ~=~ 0.056 \pm 0.005 ~{\rm keV}
\eeq
the most recent L3 results \cite{L3} find no evidence for the non-resonant
$\eta'\rta\pi^+\pi^-\c$ decay in a channel dominated by the $\r$:~
$\eta'\rta\r\c$, $\r\rta\pi^+\pi^-$. This is in contradiction with earlier
claims \cite{Bit,Ben,Abe} that the non-resonant process had been seen.

To conclude this discussion of radiative $\eta'$ and $\eta$ decays, it will
be clear that a re-analysis of the experimental data using the decay
and Dashen formulae (8),(9) should be performed to verify whether or not
$g_{G\c\c}$ is negligibly small or, if not, to measure it. We would then
have a theoretically reliable determination of all the parameters
playing a role in this sector. This would be free of any additional prior 
theoretical input based on either OZI \cite{SVeta,Seta} or the large $N_c$
expansion in the chiral Lagrangian framework \cite{KL1,KL2,Evans,Barc1,Barc2}.
At the same time, it should be recognised that
the `glue content' $\la0|Q|\eta'(\eta)\ra$ of the $\eta'$ and $\eta$
are {\it not} expressible in terms of the true, RG-invariant decay constants
$f^{a\a}$ and the relations (41) should not be used in the analysis
of $\psi\rta\eta'(\eta)\c$ data.

\section{Related experiments}

We now mention briefly a few other experiments where the theoretical
techniques described in this paper can be applied.

The first is the behaviour of the polarised photon structure function
$g_1^\c$, considered as a function of the `target photon' momentum $p$.
Standard analysis of the two-photon DIS process $e^+ e^- \rta e^+ e^- X$
reduces the problem of finding the first moment of $g_1^\c$ to a 
non-perturbative evaluation of the AVV three-point
Green function $\la0|J_{\m5}^0 J_\l^{\rm em}(p) J_\r^{\rm em}(-p)|0\ra$.
This is of course precisely the problem in evaluating $\eta'(\eta)
\rta\c\c$, except that we now want to study a range of off-shell photon
momenta. This has been carried out in refs.\cite{NSVgamma,SVgamma}. A
sum rule for $g_1^\c$ has been presented which fixes the $p\rta0$ and
asymptotically large $p$ behaviour of the first moment of $g_1^\c$ in 
terms of the electromagnetic and colour $U_A(1)$ anomaly coefficients.
For intermediate values of $p\sim m_\r$, the behaviour is shown to be
sensitive to the non-perturbative realisation of chiral symmetry breaking.
Detailed results are given in refs.\cite{NSVgamma,SVgamma}.

The theoretical techniques developed here are equally applicable to the
flavour-singlet contribution to the first moment of the polarised proton 
structure function $g_1^p$, the famous `proton spin' problem.
This is shown \cite{Vgp} to be related to a $U_A(1)$ Goldberger-Treiman 
relation, which is evaluated easily by the techniques of section 2 and
which involves an analogous new coupling $g_{GNN}$. The OZI and RG
conditions in this case are however rather different 
\cite{SVGT1,SVGT2,NSVGT3,NSVGT4}. 

More recently, it has been proposed \cite{Koch1,Koch2} that the theoretical
formalism of section 2 may also provide a way of understanding recent
data from the CLAS collaboration \cite{CLAS1,CLAS2} on the photo- 
and electro-production of $\phi$ mesons. The idea here is that one important 
contribution to the photoproduction process $\c N \rta \phi N$ may be
modelled by `glue' exchange, which in turn could be related to the
$g_{GNN}$ coupling in the $U_A(1)$ Goldberger-Treiman formula and the
$g_{G\phi\c}$ coupling extracted from $\eta'(\eta)\rta\phi\c$ as
discussed above. This is an interesting proposal which certainly
deserves to be pursued. 

Another closely related process is $\eta'(\eta)$ photoproduction.
Both $\c p\rta \eta p$ \cite{TAPS} and $\c p\rta \eta' p$ 
\cite{SAPHIR,CEBAF} have been studied in low energy experiments close
to threshold.  A theoretical discussion of these reactions,
based on an effective action incorporating the $U_A(1)$ PCAC formalism
of section 2, may be found in ref.\cite{Bass1}. 
For a recent general review of $\eta'$ physics along these lines, emphasising
the importance of the $U_A(1)$ anomaly and gluonic degrees of
freedom and incorporating an extensive survey of experiments, 
see also \cite{Bass2} and references therein.

We should also make a special mention here of the new WASA $4\pi$ detector
\cite{WASA} at CELSIUS, which will provide an important facility for precision
$\eta$  and $\eta'$ decays. This is of course the stimulus for the
present Workshop.  

Finally, and somewhat remarkably, the process $\eta'(\eta)\rta\c\c$ plays 
an important role in calculations of the strong-interaction contributions
to the anomalous magnetic moment $g_\m-2$ of the muon. This is of great current
interest because of the discrepancy between the most recent experiments
\cite{g2EXP1,g2EXP2} and theory (see e.g.~\cite{Narg2} for a recent
compilation of sub-process contributions to $g_\m-2$) and the suggestion that
this could be a signal of non-standard model physics (e.g.~\cite{Marc}).
The important contribution for our purposes is the $O(\a^3)$ correction 
arising from the Feynman diagram in which a QCD sub-diagram is linked to 
the muon line by three virtual photons, effectively a hadronic contribution
to light-by-light scattering. The dominant light pseudoscalar 
intermediate state therefore involves back-to-back $P\rta\c^*\c^*$
vertices, where of course the photons are off-shell. This is fully
discussed in ref.\cite{Kin} where it is recognised that the conventional
theoretical analysis is inadequate and a proper treatment incorporating
the gluonic anomaly along the lines discussed here is really required.
In its absence, ref.\cite{Kin} relies on using phenomenological models
and inputs from experiment. The result is that the total hadronic
light-by-light scattering contribution to the muon $g_\m-2$ is
$-79.2(15.4)\times 10^{-11}$, to be compared with the current experiment
versus theory discrepancy in $a_\m ={1\over2}(g_\m-2)$ of
$a_\m^{\rm exp} - a_\m^{\rm SM} = 333(173)\times 10^{-11}$ \cite{Narg2}.

\section{Theory}

In this section we derive the $\eta'(\eta)\rta\c\c$ decay formula using
the method of operator Green functions and generalised 1PI vertices 
developed in refs.\cite{SVeta,Seta} etc. We present the discussion in three
parts -- first the anomalous chiral Ward identities for the Green functions
of the relevant composite operators and the 1PI vertices, then the derivation
of the decay formulae in terms of 1PI vertices (which are essentially the 
couplings $g_{\eta'(\eta,G)\c\c}$~), and finally the renormalisation group 
equations.
We then comment briefly on alternative methods involving chiral Lagrangians.

\subsection{Chiral Ward Identities}
\vskip0.1cm

The anomalous chiral Ward identities for QCD with massive quarks have
been written down in the form used here in the review \cite{Szuoz}.
We refer to this article for more complete derivations and restrict ourselves
here to the most essential identities. For the moment, we omit
the electromagnetic contribution to the anomaly.

The composite operators involved in the Green functions and 1PI vertices
studied here are the currents and pseudoscalar operators 
$J_{\m5}^a$, $Q$, $\phi_5^a$ and the corresponding 
scalar $\phi^a$. We use the compact notation for the quark masses
and condensates introduced at the start of section 2.

The expressions for these operators given in section 1 define the bare 
operators. The renormalised composite operators are defined as follows:
\beqa
J_{\m5}^0 &= Z J_{\m5B}^0  ~~~~~~~~~~
J_{\m5}^{a\neq0} =  J_{\m5B}^{a\neq0} ~~~~~~~~\cr
Q &= Q_B - {1\over 2n_f}(1-Z) \pl^\m J_{\m5B}^0 ~~~~~~~~\cr
\phi_{5}^a &= Z_\phi \phi_{5B}^a ~~~~~~~~~~~~~
\phi^a = Z_\phi \phi_B^a ~~~~~~~~\cr
\eeqa
where $Z_\phi$ is the inverse of the mass renormalisation, 
$Z_\phi = Z_m^{-1}$. The non-trivial renormalisation of $J_{\m5}^0$
means that its matrix elements scale with an anomalous dimension $\c$
related to $Z$. This occurs because $J_{\m5}^0$ is not a conserved
current, due to the anomaly $Q$.
Notice in particular the mixing of the operator $Q$ with 
$\pl^\m J_{\m5}^0$ under renormalisation. As explained in \cite{ET},
this leaves the combination $(\pl^\m J_{\m5}^0 - 2n_f Q)$ 
appearing in the anomaly equation invariant under renormalisation.

The Green functions for these operators are constructed by functional
differentiation from the generating functional 
$W[V_{\m5}^a, \o, S_5^a, S^a]$, where
$V_{\m5}^a, \o, S_5^a, S^a$ are the sources for the composite
operators $J_{\m5}^a, Q, \phi_5^a, \phi^a$ respectively.
For example, the Green function $i\la0|T~Q(x)~Q(y)|0\ra$ is given by
${\d^2 W\over \d\o(x) \d\o(y)}$, which we abbreviate as $W_{\o\o}$.
This compact notation is perhaps unfamiliar, but is very convenient
for manipulating complicated expressions involving 
the chiral Ward identities.

In this functional formalism, the anomalous chiral Ward identity
is 
\beqa
\pl_{\m} W_{V_{\m5}^a} = 2n_f \d_{a0} W_{\o} + M_{ac} W_{S_5^c}~~~~~~~~~ \cr
{}~~~~~~~~~~~~~~~~~~- d_{adc} S^d W_{S_5^c} + d_{adc} S_5^d W_{S^c} \cr
\eeqa
This can be compared with eq.(15). Notice the presence of the source
variation terms on the r.h.s.
This immediately give the identities for the 2-point Green functions:
\beqa
ik_\m W_{V_{\m5}^a V_{\n5}^b} - 2n_f \d_{a0} W_{\o V_{\n5}^b} - M_{ac}
W_{S_5^c V_{\n5}^b} = 0 ~~~~~\cr
{}\cr
ik_\m W_{V_{\m5}^a \o} - 2n_f \d_{a0} W_{\o \o} - M_{ac} W_{S_5^c \o} 
= 0 ~~~~~\cr
{}\cr
ik_\m W_{V_{\m5}^a S_5^b} - 2n_f \d_{a0} W_{\o S_5^b} - M_{ac} W_{S_5^c
S_5^b} - \Phi_{ab} = 0 ~~~~~\cr
\eeqa
Combining these individual equations, we find the familiar identity 
\beqa
k_\m k_\n W_{V_{\m5}^a V_{\n5}^b} - M_{ac} \Phi_{cb} = W_{S_D^a S_D^b}
\eeqa
where $S_D^a$ is the source for the current divergence operator 
$D^a = 2n_f \d_{a0} Q + M_{ac}\phi_5^c$. 

For our purposes, we really only need the zero-momentum chiral Ward
identities. Clearly, assuming there is no massless boson coupling to the
$U_A(1)$ current, these are just
\beqa
&2n_f \d_{a0} W_{\o \o} + M_{ac} W_{S_5^c \o} = 0 ~~~~~~~~\cr
{}\cr
&2n_f \d_{a0} W_{\o S_5^b} + M_{ac} W_{S_5^c S_5^b} + \Phi_{ab} = 0 ~~~~~~~~\cr
\eeqa
which implies the following identity for the topological susceptibility,
\beq
(2n_f)^2 \chi(0) = M_{0c} W_{S_5^c S_5^d} M_{d0} + (M\Phi)_{00}
\eeq

The 1PI vertices used below are defined as functional derivatives
of a second generating functional (effective action) $\C$, constructed
from $W$ by a partial Legendre transform with respect to the fields
$Q$, $\phi_5^a$ and $\phi^a$ only ({\it not} the currents
$J_{\m5}^a$).
The resulting vertices are `1PI' w.r.t.~the propagators for these composite 
operators only. This separates off the particle poles in these 
propagators, and gives the closest identification of the field theoretic
vertices with the physical couplings such as $g_{\eta^\a \c\c}$ \cite{Szuoz}.

The basic anomalous chiral Ward identity for $\C$ follows immediately
from that for $W$:
\beqa
\pl_\m \C_{V_{\m5}^a} = 2n_f \d_{a0} Q + M_{ac} \phi_5^c ~~~~~~~~~~~~~\cr
{}\cr
{}~~~~~~~~~~~~~~~~~~~~- d_{acd}
\phi^d \C_{\phi_5^c} + d_{acd} \phi_5^d \C_{\phi^c}  \cr
\eeqa
and other identities follow simply by functional differentiation.
In particular, for the two-point vertices, we find the following
identities 
\beqa
ik_\m \C_{V_{\m5}^a V_{\n5}^b} + \Phi_{ac} \C_{\phi_5^c V_{\n5}^b} = 0 \cr
{}\cr
ik_\m \C_{V_{\m5}^a Q} - 2n_f \d_{a0} + \Phi_{ac} \C_{\phi_5^c Q} = 0 \cr
{}\cr
ik_\m \C_{V_{\m5}^a \phi_5^b} + \Phi_{ac} \C_{\phi_5^c \phi_5^b} - M_{ab}
= 0 
\eeqa
from which follows
\beqa
k_\m k_\n \C_{V_{\m5}^a V_{\n5}^b} + M_{ac} \Phi_{cb} = \Phi_{ac}
\C_{\phi_5^c \phi_5^d} \Phi_{db} 
\eeqa
At zero momentum, these reduce to
\beqa
&\Phi_{ac}\C_{\phi_5^c Q} - 2n_f \d_{a0} = 0 ~~~~~~~~~~~~~~~~~~~~~~\cr
{}\cr
&\Phi_{ac} \C_{\phi_5^c \phi_5^b} - M_{ab} = 0 ~~~~~~~~~~~~~~~~~~~~~~\cr
\eeqa
which together imply
\beq
\Phi_{ac} \C_{\phi_5^c \phi_5^d} \Phi_{db} = - (M\Phi)_{ab}
\eeq

The fact that the topological susceptibility is zero for vanishing quark mass
can be seen immediately from eq.(53). One of the simplest ways to derive
the precise form (12) or (21) is in fact to use an identity involving
$\C$. The two-point vertices are simply the inverse
of the two-point Green functions (propagators), so in the pseudoscalar sector
we have the matrix inversion formula:
\beqa
\C_{QQ} = - \Bigl(W_{\o\o} - W_{\o S_5^a} (W_{S_5 S_5})_{ab}^{-1} 
W_{S_5^b \o} \Bigr)^{-1} ~~~~~~~~~\cr
= - \Bigl(W_{\o\o} - W_{\o S_5^a} M_{ac} 
\bigl( M W_{S_5 S_5} M \bigr)_{cd}^{-1} 
M_{db} W_{S_5^b \o} \Bigr)^{-1} \cr
\eeqa
and using the identities (52) and (53) this implies
\beq
\C_{QQ}^{-1} = - \chi \Bigl(1 - (2n_f)^2 \chi (M\Phi)_{00}^{-1} \Bigr)^{-1}
\eeq
all at zero momentum.
Inverting this relation gives the important result for the topological 
susceptibility,
\beq
\chi = - \C_{QQ}^{-1} \Bigl(1 - (2n_f)^2 \C_{QQ}^{-1}
(M\Phi)_{00}^{-1} \Bigr)^{-1}
\eeq
Substituting the explicit expression for $(M\Phi)_{00}^{-1}$ (which is easily
found from the definitions above), viz.
\beq
(M\Phi)_{00}^{-1} = {1\over (2n_f)^2} \sum_q {1\over m_q \la \bar q q\ra}
\eeq
we see that (61) reproduces the general form (12) where we can
now identify the (mass-independent) non-perturbative coefficient as
\beq
A =  \C_{QQ}^{-1}
\eeq
\vskip0.1cm
We have already exploited these formulae in section 2.

\subsection{$\eta'(\eta)\rta\c\c$ from 1PI Vertices}
\vskip0.2cm

We are now ready to present the derivation of the decay formula (3) 
and generalised Dashen formula (4).  The technique relies on the
identification of the couplings $g_{\eta^\a \c\c}$ with the zero-momentum 
limit of the appropriate 1PI vertex functions introduced above. 

The starting point is the Ward identity (54) extended to include the
electromagnetic contribution to the anomaly for the axial current:
\beqa
{}\cr
\pl_\m \C_{V_{\m5}^a} = 2n_f \d_{a0} Q 
+ a_{\rm em}^a Q_{\rm em}(A) + M_{ac} \phi_5^c \cr
{}\cr
- d_{acd}\phi^d \C_{\phi_5^c} + d_{acd} \phi_5^d \C_{\phi^c} \cr 
\eeqa
$Q_{\rm em}(A)$ is just shorthand notation for ${\a\over8\pi} F_{\m\n}
\tilde F^{\m\n}$, where $F_{\m\n}$ is the field strength for the 
electromagnetic field $A_\m$. (Since we are working only to leading order in
$\a$, it is not necessary to consider $Q_{\rm em}$ as an independent
composite operator with non-trivial renormalisation.)

Differentiating twice w.r.t.~the field $A_\m$, evaluating at the VEVs,
and taking the Fourier transform, we find
\beqa
ik_\m \C_{V_{\m5}^a A^\l A^\r} = -a_{\rm em}^a {\a\over\pi} \e_{\l\s\a\b}
p_1^\a p_2^\b ~~~~~\cr
{}~~~~~~~~~~~~~~~~~~~~~~~~~~- d_{abc} \phi^c \C_{\phi_5^b A^\l A^\r} \cr
\eeqa
where $p_1,p_2$ are the momenta of the photons. Notice that the mass
term in (64) does not contribute explicitly to this formula.
From its definition as 1PI w.r.t.~the pseudoscalar fields, the vertex
$\C_{V_{\m5}^a A^\l A^\r}$ has no pole at $k^2=0$ (even in the chiral limit)
so the first term vanishes at zero momentum $k$, leaving simply
\beq
\Phi_{ab} \hat\C_{\phi_5^b A^\l A^\r}\Big|_{k=0} = a_{\rm em}^a {\a\over\pi}
\eeq
(To simplify notation, 
it will be convenient from now on to define vertices 
$\hat \C$ with the kinematical factors removed, e.g.
$\C_{\phi_5^a A^\l A^\r} = -\hat\C_{\phi_5^a A^\l A^\r} 
\e_{\l\s\a\b} p_1^\a p_2^\b$.) 

The first step in converting (66) to the decay formula (3)
is to identify the physical states $\eta^\a$. These appear as poles in the
propagator matrix for the four pseudoscalar operators $Q$, $\phi_5^a$
($a=0,3,8$). To isolate these poles, we diagonalise the propagator
matrix in this sector then normalise the three operators coupling to the 
physical states.

We therefore define the operator 
\beq
G = Q - W_{\o S_5^a} (W_{S_5 S_5})_{ab}^{-1} \phi_5^b
\eeq
so that by construction the propagators $\la G~ \phi_5^a\ra$ all vanish.
(Notice that integrations over repeated spacetime arguments are implied
in this condensed notation.) Then define operators
\beq
\eta^\a = C^{\a b} \phi_5^b
\eeq
such that the propagator matrix 
\beqa
\la \eta^\a~ \eta^\b\ra &\equiv W_{S_{\eta}^\a S_{\eta}^\b}
= C^{\a a} W_{S_5^a S_5^b} C^{T b \b} \cr
{}\cr
{}&= \left(\matrix{{-1\over k^2 - m_{\eta'}^2} &0 &0 \cr
0&{-1\over k^2 - m_{\eta}^2} &0 \cr
0 &0 &{-1\over k^2 - m_{\pi}^2}}\right) \cr
\eeqa
where $S_{\eta}^\a$ are the sources for the operators $\eta^\a$.

This change of variable affects the partial functional derivatives 
in $\hat\C_{\phi_5^a A^\l A^\r}$ in (66), which involves 
${\d\over\d \phi_5^a}$ {\it at fixed} $Q$.
In terms of the new variables $G$, $\eta^\a$ we have
\beqa
{\d\over\d \phi_5^a}\Big|_{Q} = 
{\d\eta^\a\over\d \phi_5^a}{\d\over\d \eta^\a}
+ {\d G\over\d \phi_5^a} {\d\over\d G} \cr
{}\cr
= C^{T a\a} {\d\over\d \eta^\a}
- (W_{S_5 S_5})_{ab}^{-1} W_{S_5^b \o}{\d\over\d G} \cr
\eeqa
The decay formula therefore becomes
\beqa
\Phi_{ab} C^{T b\a} ~\hat \C_{\eta^\a A^\l A^\r} -
\Phi_{ab} (W_{S_5 S_5})_{ab}^{-1} W_{S_5^b \o} ~\hat \C_{G A^\l A^\r} \cr
= a_{\rm em}^a {\a\over\pi}~~~~~~~~~~~~~~\cr
\eeqa

The decay constants are identified as
\beq
f^{a\a} = \Phi_{ab} C^{T b\a}
\eeq
In terms of the propagators, we can write (from eq.(68))
\beq
f^{a\a} (W_{S_{\eta} S_{\eta}})_{\a\b}^{-1} f^{T \b b} = 
\Phi_{ac} (W_{S_5 S_5})_{cd}^{-1} \Phi_{db}
\eeq
and so at zero momentum
\beq
f^{a\a} m_{\a\b}^2 f^{T \b b} = 
\Phi_{ac} (W_{S_5 S_5})_{cd}^{-1} \Phi_{db}
\eeq
as quoted in (28).

The remaining steps in finding (3) and (4) are an exercise in manipulating 
the zero-momentum Ward identities. First note that combining the
two identities in (52) gives
\beq
M_{ac} W_{S_5^c S_5^d} M_{db} = - (M\Phi)_{ab} + (2n_f)^2 \chi(0) 
\d_{a0} \d_{b0}
\eeq
whose $a,b=0$ component is just (53). Note that $(M\Phi)_{ab}$ is
symmetric. Also define ${\bf 1}_{00}=\d_{a0} \d_{b0}$. 
Then we can write
\beqa
\Phi_{ab} (W_{S_5 S_5})_{ab}^{-1} W_{S_5^b \o} ~~~~~~~~~~~~
~~~~~~~~~~~~~~~~~~~~~~~~~~~~\cr
= (\Phi M)_{ac} \bigl(M W_{S_5 S_5} M\bigr)_{cd}^{-1} M_{de}W_{S_5^e \o}
~~~~~~~~~~~~~~~~~~~\cr
= -2n_f (M\Phi)_{ac} \Bigl(-(M\Phi) + (2n_f)^2 \chi(0) {\bf 1}_{00}
\Bigr)_{c0}^{-1} \chi(0) \cr
= 2n_f \chi(0) \Bigl(1 - (2n_f)^2 \chi(0)
(M\Phi)_{00}^{-1}\Bigr)^{-1} \d_{a0} ~~~~~~~~~~\cr
= -2n_f \C_{QQ}^{-1}~ \d_{a0} ~~~~~~~~~~~~~~~~~~~~~~~~~~~
~~~~~~~~~~~~~~~~~\cr
\eeqa
where in the final step we have used the identification (61).
Similarly,
\beqa
\Phi_{ac} (W_{S_5 S_5})_{cd}^{-1} \Phi_{db}~~~~~~~~~~~~~~~~~~~~
~~~~~~~~~~~~~~~~~~~\cr
= (\Phi M)_{ac} \bigl(M W_{S_5 S_5} M\bigr)_{cd}^{-1} (M\Phi)_{db}
~~~~~~~~~~~~~~~~~~ \cr
= (M\Phi)_{ac} \Bigl(-(M\Phi) + (2n_f)^2 \chi(0) {\bf 1}_{00}
\Bigr)_{cd}^{-1} (M\Phi)_{db} \cr
= -(M\Phi)_{ab} + (2n_f)^2 \C_{QQ}^{-1} ~\d_{a0} \d_{b0} ~~~~~~
~~~~~~~~~~~~~~\cr
\eeqa

This establishes the required results. Substituting (76) and (77)
into (71) and (74) we find the decay formula
\beq
f^{a\a}~\hat \C_{\eta^\a A^\l A^\r}
+ 2n_f \C_{QQ}^{-1} ~\hat \C_{G A^\l A^\r}
= a_{\rm em}^a {\a\over\pi}
\eeq
where the decay constants satisfy
\beq
f^{a\a} m_{\a\b}^2 f^{T \b b} = 
-(M\Phi)_{ab} + (2n_f)^2 \C_{QQ}^{-1} ~\d_{a0} \d_{b0} 
\eeq

The final step is to identify the 1PI vertices with the couplings
defined in section 1, viz.
\beq
\hat \C_{\eta^\a A^\l A^\r} =  g_{\eta^\a \c\c}
\eeq
and similarly for $g_{G\c\c}$.
It is at this point that the central dynamical assumption is made.
In fact, eqs.(78) and (79) are exact identities, following
simply from the definitions and the zero-momentum chiral Ward identities.
To make contact with the radiative decays of the physical particles,
we must assume in particular that the 1PI vertex {\it evaluated
at $k=0$} accurately approximates the physical coupling\footnote{The 
assumption that the 1PI vertices as 
defined here can be identified at all with the decay couplings of the physical
particles rests on {\it pole dominance} -- the assumption that the dominant 
particle poles in the pseudoscalar propagator matrix are indeed those of the 
$\eta^\a$ (see eq.(69)).}, which is defined {\it on mass-shell}.
This requires that $\hat\C_{\eta^\a A^\l A^\r}$ has
only a weak momentum dependence in the range $0 \le k^2 \le m_{\eta^\a}^2$.
This is reasonable, since it is defined to be a pole-free, amputated, RG invariant
dynamical quantity. However, as in standard PCAC, the assumption is expected
to be excellent for the $\pi$ but progressively worse as the mass
of the pseudo-Goldstone bosons increases. The hope here, in common
with all attempts to include the $\eta'$ in the framework of PCAC (including
chiral Lagrangians with $1/N_c$ effects included\cite{KL1,KL2,Evans,Barc1,Barc2}, 
is that the approximation remains sufficiently good at the mass of the 
$\eta'$.

\subsection{Renormalisation Group}
\vskip0.1cm

It is important to determine the renormalisation group behaviour of all the 
quantities appearing in these formulae. Recall, for example, that the
RG behaviour was a crucial factor in the conjecture that $g_{G\c\c}$ may be 
small in the leading OZI approximation. In general, the RG equations play
a key role in understanding the physics of the $U_A(1)$ channel. We therefore
include here a brief and rather novel discussion of the RGEs for the relevant
Green functions and 1PI vertices in the functional formalism. The essential 
results were first given in ref.\cite{SVeta}, but are generalised here 
to include $SU(3)$ breaking and $\eta -\eta'$ mixing. The results are a 
straightforward extension of refs.\cite{SVeta,Szuoz} but were not explicitly
written down in \cite{Seta}.  

The fundamental RGE for the generating functional $W$ in pure QCD follows 
immediately from the definitions (48) of the renormalised composite 
operators. It is:
\beqa
\DD W = \c\Bigl(V_{\m5}^0 - {1\over2n_f}\pl_\m \theta\Bigr)W_{V_{\m5}^0}
~~~~~~~~~~~~~~~~\cr
+ \c_\phi\Bigl(S_5^a W_{S_5^a} + S^a W_{S^a}\Bigr) 
+ \ldots \cr
\eeqa
where $\DD = \Bigl(\m{\pl\over\pl\m} + \b{\pl\over\pl g} - \c_m\sum_q
m_q{\pl\over\pl m_q}\Bigr)\Big|_{V,\theta,S_5,S}$.
The notation $+\ldots$ (which is suppressed in the following equations)
refers to the additional terms of $O(k^2)$
and $O(k^4)$ which are required to produce the contact term contributions 
to the RGEs for $n$-point Green functions of composite operators. 
(This notation is omitted in the following equations, but it should be
remembered that it is implicit.)
These terms are discussed fully in refs.\cite{SRG,SVeta}, but will be omitted 
here for simplicity. They vanish at zero momentum so do not directly affect the
derivation of the decay formulae, but do have important implications for the
validity of PCAC extrapolations from zero-momentum to on-shell quantities.

The RGEs for Green functions are found simply by differentiating
eq.(81) w.r.t.~the sources. Simplifying the results using the
chiral Ward identities (50), we find a complete set of RGEs for the
2-point functions. These are:
\beqa
\DD W_{V_{\m5}^a V_{\n5}^b} = 
(\c\d_{a0} + \c\d_{0b}) W_{V_{\m5}^a V_{\n5}^b} ~~~~~~~~~~~~~~~~~~~~~~~\cr
\DD W_{V_{\m5}^a \theta} = (\c \d_{a0} + \c)W_{V_{\m5}^a \theta} 
+ \c {1\over2n_f} M_{0b} W_{V_{\mu5}^a S_5^b}  ~~~~~~\cr
\DD W_{V_{\mu5}^a S_5^b} = (\c\d_{a0} + \c_\phi) 
W_{V_{\mu5}^a S_5^b}  ~~~~~~~~~~~~~~~~~~~~~~~~~~~\cr
\DD W_{\theta \theta} = 2\c W_{\theta \theta} 
+ 2\c {1\over2n_f} M_{0b} W_{\theta S_5^b}  ~~~~~~~~~~~~~~
~~~~~~~~~\cr
\DD W_{\theta S_5^b} = (\c + \c_\phi) W_{\theta S_5^b} 
+ \c {1\over2n_f} \bigl(M_{0c} W_{S_5^c S_5^b}  + \Phi_{0b}\bigr) 
~\cr
\DD W_{S_5^a S_5^b} = 2\c_\phi W_{S_5^a S_5^b}  ~~~~~~~~~~~~~~~~~
~~~~~~~~~~~~~~~~~~~~~~~\cr
\eeqa
The pattern of cancellations which ensures the consistency of these
equations with the chiral Ward identities is quite intricate, but
may readily be checked.

At zero momentum, we can immediately use the second of eqs.(52)
to write the above RGE for $W_{\theta\theta}$ as
\beqa
\DD W_{\theta\theta} ~{\mathop=_{k\rta 0}}~ 2\c W_{\theta\theta}
~~~~~~~~~~~~~~~~~~~~~~~~~~~~~~~~\cr
-2\c{1\over(2n_f)^2}\Bigl(M_{0a}W_{S_5^a S_5^b}M_{b0} + (M\Phi)_{00}\Bigr)
~~\cr
{}\cr
= 0 ~~~~~~~~~~~~~~~~~~~~~~~~~~~~~~~~~~~~~~~~\cr
\eeqa
using (53). This shows that the zero-momentum topological susceptibility
is a RG invariant,
\beq
\DD \chi(0) = 0
\eeq
and thus
\beq
\DD A = 0
\eeq
where $A$ is the non-perturbative parameter in eq.(12), which enters into
the final decay and Dashen formulae.

Next, we need the RGE for the generating functional of the 1PI vertices.
This follows immediately from its definition and the 
RGE (81) for $W$:
\beqa
\tilde\DD\C = \c\Bigl(V_{\m5}^0 - {1\over2n_f}\C_Q\pl_\m\Bigr)
\C_{V_{\m5}^0}~~~~~~~~~~~~~~~~~~~~ \cr
- \c_\phi\Bigl(\phi_5^a \C_{\phi_5^a} + 
\phi^a \C_{\phi^a}\Bigr) +\ldots ~~~~~\cr
\eeqa
where $\tilde\DD = \Bigl(\m{\pl\over\pl\m} + \b{\pl\over\pl g} 
- \c_m\sum_q m_q{\pl\over\pl m_q}\Bigr)\Big|_{V,Q,\phi_5,\phi}$.

The RGEs for the 1PI vertices are found by differentiation, and using
the Ward identities (55) to simplify the results, we find
for the pseudoscalar sector:
\beqa
\DD\C_{QQ} = -2\c \C_{QQ}  
+ 2\c {1\over2n_f} \Bigl[\Phi_{0c} \C_{QQ} \C_{\phi_5^c Q}\Bigr] 
~~~~~~~~~~ \cr
{}\cr
\DD\C_{Q\phi_5^b} = -(\c +\c_\phi) \C_{Q\phi_5^b} ~~~~~~~~~~~~~~~~
~~~~~~~~~~~~~~~~~\cr
+\c {1\over2n_f}\Bigl[\Phi_{0c}\bigl(
\C_{QQ}\C_{\phi_5^c \phi_5^b} + \C_{Q \phi_5^c} \C_{Q\phi_5^b}\bigr)
- M_{0b}\C_{QQ} \Bigr]  \cr
{}\cr
\DD\C_{\phi_5^a\phi_5^b} = -2\c_\phi \C_{\phi_5^a\phi_5^b}~~~~~~~~~~~~~~
~~~~~~~~~~~~~~~~~~~~~~~~\cr
+ \c {1\over2n_f} \Bigl[\Phi_{0c}\C_{\phi_5^a Q}\C_{\phi_5^c\phi_5^b} 
- M_{0b}\C_{\phi_5^a Q} + ~a\leftrightarrow b~\Bigr]  ~~~~~\cr
\eeqa
Here, $\DD = \tilde\DD + \c_\phi \langle\phi^a\rangle{\d\over\d\phi^a}$.
As explained in ref.\cite{SVeta}, this is identical to the RG operator
$\DD$ defined above (acting on $W$) when the sources are set to 
zero and the fields to their VEVs.

These RGEs play two roles in the discussion.
First, they are used as consistency checks on the various
formulae we derive. Second, according to our conjecture, they provide
the clue to identifying quantities which are likely to show violations
of the OZI rule and those for which we may reasonably expect the
OZI limit to be a good approximation. This is because we can identify 
quantities which will be particularly sensitive to the $U_A(1)$ anomaly 
as those which have RGEs involving the anomalous dimension $\c$.

We now derive the RGEs for the Green functions and 1PI vertices
involved in the various expressions related to the $\eta'(\eta)\rta\c\c$
amplitude. To do this, we first need to include the electromagnetic
fields and their anomalous dimensions. As already noted in section 5.2, the
anomalous dimension for the composite operator $Q_{\rm em}(A)$ entering
the anomaly equation is of $O(\a^2)$ so can be
neglected at the order at which we are working. We denote the anomalous 
dimension corresponding to the usual electromagnetic field renormalisation
by $\c_A$.

The RGEs for the 1PI vertices $\CQA$ and $\CbA$ are easily found by
differentiating eq.(86) and simplifying using the Ward identities.
We find,
\beqa 
(\DD + 2\c_A)~\CQA =~~~~~~~~~~~~~~~~~~~~~~~~~~~~~~~\cr 
- \c \CQA +\c {1\over 2n_f} \C_{Q Q}~a_{\rm em}^0 {\a\over\pi}~\epp
~~~~~~~\cr
+ \c {1\over 2n_f} \bigl(
\Phi_{0c}\C_{Q \phi_5^c} \CQA + \Phi_{0c}\C_{Q Q} \CcA \bigr) ~~ \cr
\eeqa
and,
\beqa  
(\DD + 2\c_A)~\CbA = ~~~~~~~~~~~~~~~~~~~~~~~~~~~~~~~~~~~~~~~~\cr
- \c_\phi \CbA  +
\c {1\over 2n_f} \C_{Q \phi_5^b}~a_{\rm em}^0 {\a\over\pi}~\epp 
~~~~~~~~~~~~~~~    \cr
+ \c {1\over 2n_f} \bigl(
(\Phi_{0c}\C_{\phi_5^c \phi_5^b} - M_{0b})\CQA 
+ \C_{Q \phi_5^b}\Phi_{0c} \CcA \bigr) \cr
\eeqa
These are very similar to the corresponding equations in 
ref.\cite{SVeta}, with the obvious inclusion of the mass term 
and $SU(3)$ breaking in the VEVs and flavour mixing.

These expressions simplify remarkably at $k^2=0$. Using the
zero-momentum chiral Ward identities (57) for 
$\C_{Q \phi_5^c}$ and $\C_{\phi_5^c \phi_5^b}$, 
together with (66), we find
\beqa
&(\DD + 2\c_A) \CQA\Big|_{k=0} = 0  \cr
{}\cr
&(\DD + 2\c_A) \CbA\Big|_{k=0} = - \c_\phi \CbA(0)
\eeqa
From the latter, we immediately have
\beq
(\DD + 2\c_A) \Phi_{ab} \CbA\Big|_{k=0} = 0
\eeq
verifying the RG consistency of the basic identity (66).
 
It only remains to rewrite these results in terms of the 1PI vertices for
$\eta^\a$ and $G$. First, recall the identification of the decay constants:
\beq
f^{a\a} = \Phi_{ab} C^{Tb\a}
\eeq
where
\beq
W_{S_{\eta}^\a S_{\eta}^\b} = C^{\a a} W_{S_5^a S_5^b} C^{Tb\b}
\eeq
The l.h.s. is the propagator matrix for the physical $\eta^\a$ and is
therefore RG invariant. From the RGE for $W_{S_5^a S_5^b}$, 
we then find
\beq
\DD C^{\a a} = -\c_\phi C^{\a a}
\eeq
and it follows immediately that
\beq
\DD f^{a\a} = 0
\eeq
This confirms that the true decay constants $f^{a\a}$ are RG invariant.
Contrast this with the current matrix element definition
$\la0|J_{\m5}^a|\eta^\a\ra = ik_\m \hat f^{a\a}$, for which 
\beq
\DD \hat f^{a\a} = \c \d_{a0}\hat f^{a\a}
\eeq

Now, since $\CGA\Big|_{\eta^\a} \equiv \CQA\Big|_{\phi_5^a}$, we 
immediately deduce from eq.(90) above that
\beq
(\DD + 2\c_A) g_{G\c\c} = 0
\eeq
Finally, from eq.(70), we have
\beqa
\Phi_{ab} \CbA ~=~ f^{a\a} \CnA ~~~~~~~~~~~~~~~~~~~~~~~\cr
- \Phi_{ab} \bigl(W_{S_5 S_5}\bigr)_{ab}^{-1}
W_{S_5^b\theta} \CGA ~~\cr
\eeqa
Combining (90) with the RG identities (82), we then find after further use
of the chiral Ward identity (52) that indeed
\beq
(\DD + 2\c_A) \CnA\Big|_{k=0} = 0
\eeq
at zero momentum, i.e.
\beq
(\DD + 2\c_A) g_{\eta^\a\c\c} = 0
\eeq
as promised. In fact, if we had included the contact terms in the RG equations 
throughout, as in ref.\cite{SVeta}, we would have found at this point that the
coupling $g_{\eta^\a\c\c}(k^2)$ is actually {\it not} RG invariant for all $k$.
However, it was found in \cite{SVeta} by keeping careful track of the contact
terms that it is also RG invariant {\it on-shell}. 
This is an important point -- it is a necessary condition for the 
dynamical assumption that the on-shell couplings may be well approximated 
by their zero-momentum values, essential to the PCAC method, to be valid.

This completes our survey of the RG properties of the radiative 
$\eta'(\eta)\rta\c\c$ decay formulae confirming that, as stated in section 1, 
{\it all} the quantities appearing in the formulae are RG invariant. 
In particular, this confirms the identification of $f^{a\a}$ as the true, 
physical decay constant.

\subsection{Chiral Lagrangians, OZI  and $1/N_c$}
\vskip0.1cm

An alternative to the approach presented here is the popular method of 
chiral Lagrangians, so we include a few comments on their relation.
Chiral Lagrangians are models of low-energy QCD in which the basic fields are
chosen to parametrise the coset manifold $G/H$ (for a chiral symmetry breaking
pattern $G\rta H$) and thus lie in one-to-one correspondence with the Goldstone
bosons. The dynamics, which is determined by the isometry group of this coset 
manifold, is therefore arranged from the outset to satisfy the (zero-momentum)
chiral Ward identities. The great advantage of chiral Lagrangians is that they 
provide a systematic way of going beyond leading order in a low-momentum expansion,
higher order terms being developed by the loop expansion in this non-renormalisable
QFT \cite{GL}.

It is, however, important not to forget that chiral Lagrangians are simply 
{\it models} of QCD. They implement the chiral Ward identities in a particularly
elegant, geometric way but they still implicitly assume the same dynamical 
approximations of pole dominance (in selecting the most relevant low-energy 
states) and smoothness of momentum extrapolations that are explicit in the 
actual QCD treatment in terms of operator Green functions.

This is especially important when chiral Lagrangians are extended
\cite{KL1,KL2,Evans,Barc1,Barc2} to the non-linear $U(3)\times U(3)/U(3)$ models
incorporating the $\eta'$, which is of course not a Goldstone boson because of the
anomaly. The dynamics of these models is therefore not entirely constrained
by the geometry of the coset space but must be implemented in part by hand if they 
are to be accurate representations of true QCD. The most promising systematic
approach is to use the $1/N_c$ expansion, since at leading order in $1/N_c$
the $\eta'$ becomes a true Goldstone boson (because the anomaly is sub-leading). 
However, as we have emphasised, the leading $1/N_c$ (or the OZI
\footnote{In the text, we have preferred to refer to the OZI approximation,
rather than $1/N_c$. The OZI limit is precisely defined \cite{Vozi} as the
truncation of full QCD in which non-planar and quark-loop diagrams are
retained, but diagrams in which the external currents are attached to
distinct quark loops, so that there are purely gluonic intermediate states,
are omitted. (This last fact makes the connection with the familiar 
phenomenological form of the OZI, or Zweig, rule.) This is a more accurate 
approximation to full QCD than either the leading $1/N_c$ limit ($N_c\rta\infty$ 
at fixed $n_f$), the quenched approximation ($n_f\rta 0$ at fixed $N_c$), 
or the leading topological expansion ($N_c\rta\infty$ at fixed $n_f/N_c$).
In the OZI or leading $1/N_c$ limits, the $U_A(1)$ anomaly is absent,
there is an extra Goldstone boson, and there is no meson-glueball mixing.})
approximation, while a good approximation for some quantities, is completely 
invalid for others. In our presentation, we have pursued the consequences of the 
anomalous chiral Ward identities as far as possible without making extra
dynamical assumptions, introducing these only at the end to make contact
with the physically observed couplings and decay constants. In particular, we
have used the renormalisation group as a guide to which quantities 
we expect to have a smooth $1/N_c$ perturbation expansion and which violate 
$1/N_c$ or OZI significantly at leading order.

It would therefore be of considerable interest to make a detailed comparison
of the $1/N_c$ chiral Lagrangian predictions \cite{KL1,KL2,Barc1,Barc2} 
with those made here (and also for the closely related analysis of the 
$U_A(1)$ Goldberger-Treiman relation \cite{SVGT1,SVGT2,NSVGT3,NSVGT4}
and its link with the `proton spin' structure function $g_1^p$).
Since the fundamental anomalous symmetry and dynamical assumptions should 
be the same, it would be interesting to see how these are realised in these two, 
in principle equivalent, approaches.

\section{Epilogue}

In this paper, we have reviewed the theory and phenomenology of the
radiative $\eta'(\eta)\rta\c\c$ decays, together with closely related
processes such as $\eta'(\eta)\rta V\c$, $\eta'(\eta)\rta\pi\pi\c$, 
$\psi\rta\eta'(\eta)\c$, etc.
The theory of these decays is indeed a tale of two anomalies: first, the
electromagnetic $U_A(1)$ anomaly, which was spectacularly successful
historically in explaining the otherwise mysterious $\pi\rta\c\c$ decay;
second, the gluonic $U_A(1)$ anomaly, which makes the physics of the
flavour-singlet $0^-$ channel in QCD so subtle and interesting.

Indeed, it is the role of the gluonic $U_A(1)$ anomaly that makes the
$\eta'$ and its decays worth studying. It is therefore disappointing that 
this new physics is so often obscured by phenomenological analyses which
try to fit the data into the straightjacket of decay formulae written down
in naive analogy with $\pi\rta\c\c$, without taking the implications of
the gluonic anomaly fully into account. The purpose of this paper is to urge a
fresh phenomenological look at $\eta'$ physics, treating both the electromagnetic
and colour $U_A(1)$ anomalies in a complete and theoretically self-consistent
manner.

The goal, going beyond mere confirmation of the well-established physics
of pseudo-Goldstone bosons and their interactions, is to gain new 
phenomenological insight into the rich and fascinating subject of
gluon topology in QCD.

\section{Acknowledgments}

This work was partly performed while I was a visiting Directeur de Recherche
at the Laboratoire de Physique Math\'ematique et Th\'eorique, 
Universit\'e de Montpellier II. I am grateful to everyone at LPM
for their hospitality and the CNRS for financial support. This research was
also supported in part by PPARC grant PPA/G/O/2000/00448.

\end{document}